%% file: mctba3.tex
\input phyzzx
\input mydef
\date{June, 2005}
\titlepage
\title{Monte--Carlo Thermodynamic Bethe Ansatz}
\author{Doron Gepner}
\line{\hfill Department of Particle Physics\hfill}
\line{\hfill Weizmann Institute\hfill}
\line{\hfill Rehovot 76100, Israel\hfill}
\abstract
We introduce a Monte--Carlo simulation approach to thermodynamic Bethe
ansatz (TBA). We exemplify the method on one particle integrable
models, which include a free boson and a free fermions systems along
with the scaling Lee--Yang model (SLYM). It is confirmed that the 
central charges and energies are correct to a very good precision,
typically 0.1\%  or so. The advantage of the method is that it enables
the calculation of all the dimensions and even the particular
partition function.
\endpage

Conformal field theory in two dimensions and its perturbed integrable models
have attracted considerable attention, since the work \REF\BPZ{A.A. Belavin, 
A.M. Polyakov and
A.B. Zamolodchikov, Nucl. Phys. B 241 (1984) 333.}\r\BPZ. A great deal of interest
stems from the work of Al. Zamolodchikov who considered first the thermalization
of integrable systems 
\REF\Zam{Al.B. Zamolodchikov, Nuclear Physics B342 (1990) 695.} (see ref. \r\Zam\
and ref. therein). 
See also \REF\TBA{T.R. Klassen and E. Melzer, Nucl. Phys. B338 (1990) 485; Nucl. Phys. B350
(1991) 635; Al. B. Zamolodchikov, Nucl. Phys. B 358 (1991) 497; Phys. Lett. B253
(1991) 391.}\r\TBA, and for further developments see
\REF\TBAtwo{F. Ravanini, Phys. Lett. B282 (1992) 73;
R. Tateo, Int. J. Mod. Phys. A10 (1995) 1357;
P. Dorey, R. Tateo and K.E. Thompson, Nucl. Phys. B470 (1996) 317;
M.J. Martins, Phys. Lett. B277 (1992) 301; Nucl. Phys. B394 (1993) 339;
V. Fateev and Al. Zamolodchikov, Phys. Lett. B271 (1991) 91;
E. Quattrini, F. Ravanini and R. Tateo, hep-th/9311116 (Cargese workshop, ``New 
Developments in String Theory, Conformal
Models and Topological Field Theory", (1993);
P. Dorey, J.L. Miramontes, Nucl. Phys. B697 (2004) 405;
P. Dorey, C. Dunning and R. Tateo, Nucl. Phys. B578 (2000) 699;
F. Woynarovich, Cond--Mat/0402129, July (2004).}\r\TBAtwo.
Here we offer an approach to thermodynamic Bethe ansatz based
on a Monte--Carlo simulation.
\par
We consider a a collection of particles moving in a box of length $l$.
For simplicity we assume only one particle specie in a purely inelastic
scattering matrix, $S$, and also assume periodic boundary condition on the box.

We assume that the particles are at the locations $x_1,x_2,\ldots x_n$ and that 
the distances $x_i-x_j$ are large, so that we can use the scattering matrix to 
compute the relative phase. We denote the scattering amplitude by $S(\beta)$
where $\beta$ is the relative rapidity, $p=m\sinh\beta$ is the momentum, 
$E=m\cosh\beta$ is the energy, and where $m$ is the mass of the particle. 

The Bethe ansatz assumes that the wave function is
$$\psi(x_1,x_2,\ldots,x_n)=\prod_r e^{i x_r p_r} \theta(x_1<x_2<\ldots<x_n)+\ldots,\e$$
where the theta ensures that $x_1<x_2<\ldots<x_n$ and the dots correspond to other 
arrangements of the coordinates, picking up a factor of $S(\beta_i-\beta_j)$ every time $x_i$
crosses $x_j$. For example for two particles, the wave function is
$$\psi(x_1,x_2)=e^{i x_1 p_1+ix_2 p_2}\left[\theta(x_2-x_1)+S(\beta_1-\beta_2) 
\theta(x_1-x_2)\right].\e$$ 

The amplitude in a unitary theory can be written as,
$$S(\beta)=e^{i \delta(\beta)},\e $$
where  $\delta(\beta)$ is a real phase.

The basic equation follows from taking a particle all around the box,
picking up a phase each time we exchange two particles location and a phase
from the wave function factor $e^{i p_i x_i}$.
It is 
$$e^{ip_i l}\prod_{j \atop j\neq i} S(\beta_i-\beta_j)=1.\e$$
Taking a log of this equation we arrive at the Bethe equation,
$$m l \sinh \beta_i+\sum_{j \atop j\neq i} \delta(\beta_i-\beta_j)=2\pi n_i,\e$$
where $i$ and $j$ take the values $1$ to $n$, and
$n$ is the number of particles.
$n_i$ are some integers which determine the energy 
levels. \foot{Denoting the wave function for these $n_i$'s by $\psi_{n_1,n_2,\ldots ,n_m}(x_i)$,
the actual wave function has to take into account that these are identical particles.
Thus, the full wave function assumes the form:
$$\twidle\psi(x_i)=\sum_p h^{s(p)} \psi_{p(n_1),p(n_2),\ldots, p(n_m)}(x_i),$$
where $p$ is any permutation of $m$ objects, and $s(p)$ is $0$ ($1$) if $p$ is an 
even (odd) permutation,
respectively, and $h=1$ ($h=-1$) for Bose (Fermi) statistics, respectively. It
follows that the wave function is defined for $n_1\leq n_2\leq \ldots\leq n_m$ for 
Bose statistics, and, $n_1<n_2<\ldots <n_m$ for Fermi statistics.}

The energy of this collection of particles is given by
$$\bar E=r E=mr \sum_{i=1}^n \cosh\beta_i,\e$$
where $r$ is the inverse temperature,
$$r={1\over kT},\e$$
where $T$ is the temperature, and $k$ is Boltzmann constant. We redefined 
the energy by multiplying it by $r$ and denoted by $\bar E$.

We define a grand canonical partition function by summing over all the particle numbers,
$m$, without a chemical potential, $\mu=0$, and summing over the energy levels,
$n_i$, taking into account that all the particles are identical.
These means that we have to limit the sum on $n_1\leq n_2\leq\ldots\leq n_m$.
The partition function then becomes,
$$Z(m,r,l)=\sum_{n_i,m \atop n_1\leq n_2\leq\ldots\leq n_m}
 \exp[-\bar E(n_i)]. \e $$
In the case of Fermi statistics, when the $n_i$ are all diferent
we may write the partition function as,
$$Z(m,r,l)=\sum_{n_i, m\atop n_i\neq n_j} {1\over m!}
\exp[-\bar E(n_i)].\e$$ 
The average 
energy is given, as usual, by
$$\bar E (m,r,l)=\langle \bar E \rangle =-r{\partial\over \partial r} \log Z(m,r,l).\e$$

The conformal limit is obtained by taking the mass to zero, $m\rarrow 0$.
In this limit the partition function becomes only a function of the
ratio $\tau=r/l$. 
$$\lim_{m\rarrow 0} Z(m,r,l)=Z(r/l),\e$$
and we denote by $\bar E(r/l)=\lim_{m\rarrow0} \bar E(m,r,l)$ the average energy in this limit.
We expect the partition function to be equal  
to the path integral on the torus with a modulus $i\tau$ of the corresponding
conformal field theory. Denote by $\cal H$ the Hilbert space of this theory.
Then the following correspondence emerges,
$$Z(r/l)=\Tr_{\cal H} e^{-2 \pi r(L_0+\bar L_0)/l},\e$$
where $L_0$ and $\bar L_0$ are the left and right moments of the stress 
energy tensor, i.e., the Hamiltonians of the system, whose eigenvalues are 
the conformal 
dimensions. The relation, eq. (12), can be used to evaluate all the dimension of 
the CFT and even the particular modular invariant used, by evaluating the
partition function.

We can now use modular invariance \REF\mod{J.L. Cardy, Nucl. Phys. B270 [FS16] (1986) 186.}\REF
\GW{D. Gepner and E. Witten, Nucl. Phys. B278 (1986) 493.}\r{\mod,\GW}.
This means that the function,
$$f(r/l)=Z(r/l) e^{2 \pi r c/(12 l)},\e$$
where $c$ is the central charge, is invariant under $l$ and $r$ replaced,
$$f(l/r)=f(r/l).\e$$
By deriving the log of this equation with respect to $r$, 
we get a relation among energies,
$$\bar E(r/l)+\bar E(l/r)={2 \pi c\over 12} \left({l\over r}+{r\over l}\right).\e$$
This is a very useful relation for evaluating the central charge.
An exception to this relation occurs for free bosons where: 1) There is an infinite
factor in $Z$ due to the zero mode, $n_i=0$, which we can ignore (or eliminate 
by giving a chemical potential $\mu$, for the $n_i=0$ mode. 2) $f$ has to be multiplied
by 
$$(l/r)^{1/2} f(r/l),\e$$
to make it modular invariant. This changes the relation eq. (15), to be
$$\bar E(r/l)+\bar E(l/r)+1={2 \pi c\over 12} \left({l\over r}+{r\over l}\right),\e$$
where the central charge $c=1$ for one free boson.

We start by studying some simple examples, which we take to be systems of
one free boson or one free fermion.

For a free boson the different modes are independent and the partition function
for the system is readily evaluated to be,
$$Z(r/l)=\prod_{m=-\infty\atop m\neq0}^\infty \sum_{s_m=0}^\infty e^{-2 \pi r |m| s_m/l}=\prod_{n=1} ^\infty (1-q^n)^{-2},\e$$
where $s_m$ is the number of particles at the $m$th energy level and we defined $q$
to be
$$q=e^{-2 \pi r/l}.\e$$
We ignored the zero mode which is a constant (infinite) factor, which can be
controlled by adding a chemical potential, as explained above.

We see immediately, that eq. (12) is indeed obeyed, and that $Z$ is exactly
the partition function of one free boson, on a torus with a modulus $i\tau$.
We can also verify that equation (17) holds giving the central charge $c=1$ as
expected.

Let us consider now a system of one free fermion. We assume that $\delta(\beta)=0$
and that the boundary condition is anti-periodic. We take a Fermi statistics,
i.e., only one particle is allowed for each energy level. This implies that the Bethe equation,
eq. (5), becomes,
$$m l \sinh\beta_i=2 \pi (n_i+1/2),\e$$
and the partition function becomes,
$$Z(r/l)=\prod_{m=0}^\infty (1+q^{m+{1\over2}})^2,\e$$
where $q$ is again given by eq. (19).

Again, this is precisely the partition function for a free fermion in the
Neveu--Schwarz sector, confirming eq. (12). We can verify that eq. (15) holds,
giving the correct central charge, $c=1/2$.

The Ramond sector is obtained by taking the trivial scattering matrix $S=1$, periodic
boundary conditions and Fermi statistics. The partition function can be easily evaluated
using eqs. (5,9), and it is
$$Z(r/l)=2\prod_{n=1}^\infty (1+q^n)^2.\e$$
We see that, indeed, it is identical to the free fermion CFT partition function,
which is
$$Z_{\rm cft}(r/l)=2 q^{1/12} \prod_{n=1}^\infty (1+q^n)^2.\e$$
We note however, that the overall factor is different. Thus we define 
$$\bar\chi_{\rm R}=q^{-\Delta+c/24} \chi_R,\e$$
where $\Delta$ is the dimension of the primary field corresponding to this block,
and $\chi_{\rm R}$ is the usual Ramond character. We then see that the correspondence
with the CFT is,
$$Z(r/l)=2 |\bar\chi_R|^2.\e$$

We conclude that two differences arise, in general, in eq. (12):
1) Each block in the CFT occurs for a different boundary conditions. 2) The
overall factor eq. (24) has to be eliminated, making the character into a ``q-series",
i.e., only integer non--negative powers of $q$ are allowed in the character.
 
Now, for an interacting system it is quite difficult, in general, to compute this partition
function. For example, the Lee--Yang theory $M(2,5)$, which is the 
perturbed minimal model with $p=2$
and $q=5$ \r\BPZ. This theory has only one particle in the spectrum, particle $A$ with a mass $m$
and the scattering amplitude is \REF\Cardy{J.L. Cardy and G. Mussardo, Santa Barbara preprint 
93106, 1989.}\r\Cardy,
$$S_{AA}(\beta)={\sinh\beta+i\sin(\pi/3) \over \sinh\beta-i\sin(\pi/3)}.\e$$ 

Thus, we resort to a numerical
algorithm for evaluating the partition function $Z(m,r,l)$. There are two stages in this algorithm, which are basically,
1) Solving the basic equation (5). 2) Simulating the grand canonical ensemble eq. (8).

To solve the equation we must calculate $\bar E(n_i)$ for any set of levels,
$n_i$. This we do by taking the initial guess $\beta_i^{0}=0$ and
solving eq. (5) by iterations,
$$\beta_i^s= \sinh^{-1} \left [{ 2\pi n_i-\sum_{j\neq i} \delta(\beta_i^{s-1}-
\beta_j^{s-1} ) \over m l}\right ],\e$$ 
where $s$ denotes the $s$th iteration. Usually $5$ or so iterations are enough to
get $\beta_i$ with sufficient precision ($10^{-6}$ or so).

The second stage we use is to make a Monte--Carlo simulation of the system using a Metropolis 
type algorithm, in order
to find the average energy, $\bar E(m,r,l)$. We use three types of steps,
preserving the detailed balance to get the correct population of particles.

For the first step, we change any of the $n_s$'s for all the $s$ by one 
$n_s\rarrow n_s\pm 1$,
$$n_1,n_2,\ldots , n_m\longrightarrow n_1,n_2,\ldots,n_{s-1},n_s\pm 1,n_{s+1},\ldots,n_m.$$
If
the energy of the new configuration is less, we accept the new configuration. If the 
energy is greater, we accept
it with the probability 
$$\exp[\bar E_{\rm old}-\bar E_{\rm new}],\e$$
which ensures the detailed balance distribution of probability $\exp[-\bar E]$. 
This step allows us to move in momentum space, 
corresponding to a thermalized random walk in the $n$ lattice,
covering all the possible values
of $n_s$ (ergodicity).

The second and third steps allow for changing the numbers of particles. i.e.,
we either insert a new level somewhere, $n_s=0$ for some $s$, or eliminating a 
level which is zero $n_r=0$. The adding of a particle corresponds to the step:
$$n_1,n_2,\ldots , n_m \longrightarrow n_1,n_2,\ldots,n_{s-1},0,n_{s+1},\ldots , n_m \e$$
and the elimination of a particle corresponds to the step
$$n_1,n_2,\ldots,n_{s-1},0,n_{s+1},\ldots , n_m\longrightarrow n_1,n_2,\ldots , n_m. \e$$

Again we take the new configuration if its energy is lower.
If the new configuration is higher in energy,
we take the new configuration with the probability eq. (28). This
allows the change of the number of particles and the population of energy levels 
is according to 
the 
grand canonical distribution eq. (8). The relation eq. (28) 
ensures the correct detailed balance. We use configurations of $n_s=0$
as ``gateways" to change the number of particles, deleting or adding
only zero energy levels.

Actually, the algorithm above does not take into account
the ordering of the energy levels,
$n_1\leq n_2\leq \ldots \leq n_m$. This we do in a different way for 
bosons and fermions. For fermions we add a particle with the 
probability $\exp[-\Delta E]/(m+1)$ if the energy is higher,
and $1$ if it is lower. We delete a particle with the energy
$\exp[-\Delta E]$ if the energy is higher and $1/(m+1)$ if
it is lower. If the $n_i$ already contains some $n_i=0$ we do not add a particle,
and leave the configuration unchanged. Likewise if the configuration does not
contain $0$ we do not delete a particle.

For free fermions this algorithm can be simplified since adding 
a particle always raises the energy, and deleting it always
lowers it. So, we add a particle with the probability 
$\exp[-\Delta E]/(m+1)$ and remove it with the probability $1$.

For bosons we use the same algorithm but allow only the configurations
which are ordered: $n_1\leq n_2\leq\ldots \leq n_m$ discarding any
other new configurations. We change $n_i\rarrow n_i\pm 1$ at some
location. If the new configuration is not ordered, we do not make the change.
This guarantees detailed balance. 

This allows us to calculate the energies, eq. (10),
preserving correctly the detailed balance.

The easiest systems to simulate are, of course, the free fermions and free boson
systems. For a free fermion system we took $\delta(\beta)=0$ along
with antiperiodic boundary condition and Fermi statistics.
For $r/l\leq 1$ we took $10000000$ sweeps for each temperature,
and for $r/l\geq 1$ we took $400000000$ sweeps.
The results are listed in tables (1) and (2). Here $\bar E_{\rm mc}$ 
and $N_{\rm mc}$
are the average energy and average particle numbers, as found in the
simulation. We define the ``partial" central charge as:
$$c_{\rm mc}(r/l)={12 \bar E_{\rm mc}(r/l)\over 2\pi
(l/r+r/l)},\e$$
and the central charge is given by, using eq. (15),
$$c_{\rm mc}(r/l)+c_{\rm mc}(l/r)=c,\e$$
where $c$ is the Monte--Carlo value of the central charge of 
the theory.
For example we can calculate, from tables (1-2),
$$c_{\rm mc}(0.5)+c_{\rm mc} (2)=0.481836+0.0178107=0.499647,\e$$
which is impressively close to the actual central charge $c=0.5$,
in accordance with eq. (15).

$\bar E_{\rm calc}$ is the energy calculated from eq. (10).
Here the average number of particles $N_{\rm calc}$ is calculated by
introducing a chemical potential $\mu$,
$$Z(r/l,\mu)=\sum_{n_i,m \atop n_1\leq n_2\leq\ldots\leq n_m} \exp[-\bar E
(n_i)-\mu m],\e$$
for Bose statistics. For Fermi statistics we have,
$$Z(r/l,\mu)=\sum_{n_i,m \atop n_1<n_2<\ldots< n_m} \exp[-\bar E(n_i)-\mu m]=
\sum_{n_i,m \atop n_i\neq n_j} {1\over m!} \exp[-\bar E(n_i)-\mu m].\e$$ 
For a free fermion $Z$ assumes the form,
$$Z(r/l,\mu)=\prod_{n=0}^\infty \left(1+\exp[-2 \pi r (n+1/2)/l-\mu]
\right)^2,\e$$
and for a free boson it is
$$Z(r/l,\mu)=\prod_{n=1}^\infty \left( 1-\exp[-2 \pi r n/l-\mu] 
\right)^{-2}.\e$$
The energy is given as in eq. (10), 
$$\bar E_{\rm calc}=-r {\partial \over\partial r} \log Z(r/l,0),\e $$
We calculate the derivative with respect to $\mu$ to get 
the average
number of particles,
$$N_{\rm calc}=-\lim_{\mu\rarrow0} {\partial\over\partial\mu} 
\log Z(r/l,\mu).\e $$

From tables (1-2), we see that across the range of temperatures
the average energy and average $N$ are typically only $0.1\%$
off the calculated numbers, which is a very reasonable correspondence.
This demonstrates the very agreeable efficacy of the Monte--Carlo
approach. 

In tables (3--4) we list the results for a free boson system,
for $r/l\leq1$ (table (3)) and $r/l\geq1$ (table (4)). 
We take a chemical potential for the zero modes, $\mu=0.5$. At this value
only a few $n_i=0$ are created. Also we make $100000000$ sweeps of the Monte--Carlo.
Again,
the correspondence between the Monte--Carlo results and the calculated
ones is very good, around $0.1\%$. We defined here $c_{\rm mc}$ as
$$c_{\rm mc}(r/l)={12 [E(r/l)+1/2]\over 2\pi (l/r+r/l)}.\e$$
It follows from eq. (17) that the Monte Carlo value of the central
charge is, again,
$$c_{\rm mc}(r/l)+c_{\rm mc}(l/r)=c_{\rm mc}.\e$$
For example, from tables (3--4),
$$c_{\rm mc}(0.8)+c_{\rm mc}(1.25)=0.528855+0.471513=1.000368,\e$$
which is very close to the theoretical value $c=1$. 

The zero modes of the bosonic system, i.e., for $n_i=0$ have to be treated 
carefully, since they have zero energy. First, we introduce a chemical potential,
{\bf only} for the zero modes, in the Monte--Carlo simulation. Second, 
in calculating $N_{\rm mc}$ we count only
the number of the non--zero modes, in order to correspond with eq. (37), the partition 
function. We see from tables (3--4) that this leads to a very good agreement
with the theoretical results.

We now turn our attention to the scaling Lee Yang model, which is the perturbed 
minimal model $M(2,5)$. We first need to establish some facts about the minimal models,
in general \r\BPZ. The models are labeled by two integers $p$ and $\pr p$ assumed
to be strange to each other, and labeled by $M(p,\pr p)$. The central charge is
$$c=1-{6(p-\pr p)^2\over p \pr p}.\e$$
The fields are given by $n=1,2,\ldots ,p-1$ and $m=1,2,\ldots, \pr p-1$, denoted by
$\phi_{n,m}$ whose dimensions
are
$$\Delta_{n,m}={(n \pr p-m p)^2-(p-\pr p)^2 \over 4 p \pr p}.\e$$
The field $\phi_{n,m}$ is identical to the field $\phi_{p-n,\pr p-m}$.
These models are unitary only for $\pr p=p+1$
\REF\Friedan{D. Friedan, Z. Qiu and S. Shenker, Phys. Rev. Lett. 52 (1984) 1575.}\r\Friedan.
In particular, the model $M(2,5)$
is not unitary. However, it is only ``weakly" non--unitary, since the scattering
matrix $S_{AA}$, eq. (26) is unitary and also, as we shall see, the model is modular invariant.
This enables us to calculate the partition function as if it was a unitary theory.

The character of the field $\phi_{m,n}$ is defined by
$$\chi_{n,m}=\Tr_{{\cal H}_{n,m}} \exp[2 \pi i \tau (L_0-c/24)].\e$$
We define the classical theta functions at the level $p$ by
\REF\Kac{V. G. Kac,``Infinite Dimensional Lie Algebras", Cambridge University Press, Third
edition (1990).}\r\Kac,
$$\Theta_{n,p}(\tau)=\sum_{j={n\over 2 p}+{\rm integer}} e^{2 \pi i p j^2 \tau},\e$$
where $n$ is defined modulo $2 p$. The theta functions are modular forms,
transforming by the modular transformation $\tau\rarrow -1/\tau$ as \r\Kac
$$\Theta_{n,p}(-1/\tau)=(-i\tau)^{-1/2} \sum_{\bar n} S_{n,\bar n} \Theta_{\bar n,p}(\tau)
,\e$$
where the sum is over $\bar n$ modulo $2 p$ and the matrix $S$ is given by
$$S_{n,\bar n}={1\over\sqrt{2 p}} e^{-\pi i n\bar n/p}.\e$$ 
The charaters of the minimal model,
$M(p,\pr p)$ can then be seen to be,
$$\chi_{n,m}(\tau)={\Theta_{n_-,p \pr p}(\tau)-\Theta_{n_+,p\pr p}(\tau) \over
\eta(\tau)},\e$$
where we defined,
$$n_\pm=n \pr p\pm m p,\e$$
and where the Dedekind eta function is defined by,
$$\eta(\tau)=q^{1/24} \prod_{n=1}^\infty (1-q^n).\e$$

It is immediately seen that these characters give the correct central charge and dimensions,
eq. (43-44). Moreover, they give the null vector for the field $\phi_{n,m}$ at the levels 
$\Delta=\Delta_{n,m}+m n$ and $\Delta=\Delta_{n,m}+(p-n) (\pr p -m)$, in accordance with
\r\BPZ.

The characters, so defined, are modular functions of a subgroup of the modular group.
They are seen to transform according to,
$$\chi_{n,m}(-1/\tau)=\sum_{\bar n,\bar m} W_{n,m,\bar n,\bar m} \chi_{\bar n,\bar m}(\tau),\e$$
where the matrix $W$ is can be seen, from eq. (48), to be
$$W_{n,m,\bar n,\bar m}=\sqrt{{8\over p\pr p}}
(-1)^{n \bar m+m\bar n} \sin\left({\pi m\bar m p \over \pr p}\right) \sin
\left({\pi n\bar n \pr p\over p}\right).\e$$

It is immediately seen that the theory $M(p,\pr p)$ is equivalent, at the level of the
modular matrix, to $SU(2)\times SU(2)$ at the ``pseudo" levels $p/\pr p$ and
$\pr p/p$ \REF\Long{D. Gepner, Caltech preprint CALT-68-1825, Hep-th 9211100 (1992).}
\REF\Imb{C. Imbimbo, Nucl. Phys. B384 (1992) 484.}\r{\Long,\Imb}, 
$$M(p,\pr p)\approx SU(2)_{p/\pr p}\times SU(2)_{\pr p/p}.\e$$
The factor $(-1)^{n\bar m+m\bar n}$ in equation (53) can be ignored since
the left movers and the right movers differ by an even integer. Here, twice the $SU(2)$ 
isospin is
given by $j=n-1$, $l=m-1$, $\bar j=\bar n -1$, $\bar l=\bar m-1$, 
and the level is $k=p-2$, $\pr k=\pr p-2$. In this notation, the modular matrix of 
$SU(2)_{\pr p/p}$ is given by \r{\Kac,\GW,\Long,\Imb}
$$S_{j,\bar j}=\sqrt{2\over k+2} \sin\left(\pr p {\pi (j+1)(\bar j+1)\over k+2} \right),\e$$
and, indeed, eq. (53) is exactly a product of two such factors, for $p/\pr p$ and 
for $\pr p/p$.
For the modular transformation $T$, which is generated by $\tau\rarrow \tau+1$, we have
$$T_{n,m}=e^{2\pi i(\Delta_{n,m}-c/24)}=e^{2\pi i n^2 \pr p/(4 p)} e^{2\pi i m^2 p/(4 \pr p)}
(-1)^{mn} e^{-2 \pi i \twidle c/24}.\e$$
The factor $(-1)^{mn}$ is irrelevant since it cancels between the right and left 
movers, as they differ by an even number, and so is the factor with the central charge.
The other factors correspond exactly to the dimensions of $SU(2)_{p/\pr p}$ and 
$SU(2)_{\pr p/p}$,
$$\Delta_j=\pr p {j(j+2)\over 4(k+2)},\e$$
and
$$\Delta_l=p {l(l+2)\over 4(\pr k+2)}.\e$$
We infer that, also, the modular matrix $T$ is a product of two $SU(2)$'s at the pseudo
levels $p$ and $\pr p$. (At the level of the dimensions, the correspondence with 
$SU(2)\times SU(2)$ was noted already in \REF\Imbim{C. Imbimbo, Private communication.}
\r\Imbim).

We conclude that the general modular invariant partition function is given by,
$$Z(\tau)={1\over 2} \sum_{n,m,\pr n,\pr m} N_{n,\pr n} K_{m,\pr m} \chi_{n,m}(\tau) \chi_{
\pr n,\pr m}(\tau)^*,\e$$
where the factor of $1/2$ accounts for the field identifications, and 
where $N$ and $K$ are any modular invariants of $SU(2)$ at the levels $k=p-2$ and 
$\pr k=\pr p-2$,
respectively, which are in relation with the simply laced Lie algebras
of types ADE \REF\Cap{D. Gepner, NPB 287 (1987)
111; A. Cappelli, C. Itzykson and J.B. Zuber, NPB 280 [FS16] (1987) 455.}\r{\GW,\Cap}.
For a proof see \REF\Proof{D. Gepner and Z. Qiu, Nucl. Phys. B285 [FS19] (1987) 423;
A. Cappelli, C. Itzykson and J. B. Zuber, Commun. Math. Phys. 113 (1987) 1.}\r\Proof.
This solves the problem of classifying the acceptable partition functions.

The modular transformations also imply that the fusion rules of the model $M(p,\pr p)$
are the same as a product of two $SU(2)$'s,
$$\phi_{n,m}\times \phi_{\bar n,\bar m}=\sum_{q,t} f^p_{n,\bar n,q} f^{\pr p}_{m,\bar m,t}
\phi_{q,t},\e$$
where $f^p_{n,\bar n,q}$ is the fusion rule of $SU(2)$ at the level $k=p-2$ according to the
``depth rule" \r\GW. We have that $f^p_{n,\bar n,q}$ is equal to one if: 
$$n+\bar n+q=1{\ \rm mod\ }2, \qquad q\geq |n-\bar n|+1,\qquad q\leq\min(n+\bar n-1, 2p-n-\bar n-1)
,\e$$
and is zero otherwise. These fusion rules can be seen to be in accordance with the 
Verlinde formula \REF\Verlinde{E.P. Verlinde, Nucl. Phys. B300 (1988) 360.}\r\Verlinde.
When we use the $S$ matrix eq. (53) we recover the correct fusion rules, eqs. (60--61). 
This implies that these are fully consistent conformal data for a {\rm unitary} theory,
obeying all the axioms of unitary conformal data.

Actually, the model $M(p,\pr p)$, for $p-\pr p$ odd, 
has exactly the same 
conformal data
as the unitary model 
$$L=SU(2)_1\times SU(2)_{(\pr p/p)-1} \times SU(2)_{(p/\pr p)-1},\e$$
where we include in the chiral algebra the field $\rho=[1]\times [p-2] \times [\pr p-2]$, and
where each of the numbers correspond to twice the isospin of $SU(2)$. The central charge
of the model is calculated to be,
$$c=1+3 (\pr p/p-1)(p-2)+3(p/\pr p-1) (\pr p-2)=1-{6(p-\pr p)^2\over p\pr p},\e$$
i.e., exactly as the central charge of $M(p,\pr p)$, eq. (43).
The field $\rho$ has integer dimension, for $p-\pr p$ odd, and so we can include it in the
chiral algebra. Also, the consistency of the pseudo conformal field theory, i.e., at
the level $p-\pr p$ requires that $p-\pr p$ is odd \r\Long.
The extended 
field implies: 1) The field identifications, $\phi_{n,m}=\phi_{p-n,\pr p-m}$, by multiplying
any of the fields with this field.
2) The total isospin has to be an integer, for locality with respect to this field:
$[s]\times [j] \times [l]$ obeys $s+j+l=0 {\ \rm mod\ } 2$.
It can be easily verified that the modular properties of the theory 
$L$, with this extended field, are exactly the same as $M(p,\pr p)$ theory, eqs. (53,56).
In particular, this implies the fusion rules, eqs. (60-61), using, for example,
the Verlinde formula \r\Verlinde. 
For $\pr p=p+1$
we recover the conformal data of the coset model $SU(2)_1\times SU(2)_p/SU(2)_{p+1}$,
as is well known \REF\GKO{P. Goddard, A. Kent and D. Olive, Commun. Math. Phys.
103 (1986) 105.}\r\GKO.

A similar correspondence holds when $p$ and $\pr p$ are both odd. In this case we define
the model
$$M\approx SU(2)_{(p+\pr p)/p}/Z_2\times SU(2)_{p/\pr p},\e $$
where $SU(2)_k/Z_2$ stands for the conformal field theory of integer isospin 
representations of $SU(2)$, see
\r\Long. It can be immediately verified that the central charge is correct, giving eq. (43),
up to an integer, and that the modular transformations are identical with eqs. (53,56).
We conclude that for all the minimal models, their conformal data can be realized by 
full fledged unitary conformal field theories.  

For the model $M(2,5)$, the SLYM, we have two fields $\phi_{1,1}=1$, which is the 
unit operator, and $\phi=\phi_{1,2}$ which has the dimension $\Delta_{1,2}=-1/5$.
According to eq. (59), only the diagonal modular invariant is allowed in this case.

The fusion rules are easily read from eqs. (60--61) and are given by,
$$\phi^2=1+\phi.\e$$
These are the same fusion rules as $(G_2)_1$ and $SU(2)_3/Z_2$.
The later is the conformal theory consisting of the integer isospin representations of 
$SU(2)_3$, see \r\Long.
However, the modular
matrix of $M(2,5)$ is at pseudo level $3$ of $SU(2)_3/Z_2$ and so is different from 
these cases (which are at the pseudo levels $\pm 1$).

It is convenient to define the characters,
$$\bar \chi_{m,n}(\tau)=q^{-\Delta_{m,n}+c/24} \chi_{m,n}(\tau),\e$$
so that $\bar\chi$ contains only integer powers of $q$ (a ``q-series") starting at $1$.
The square of this object is what we actually compute in the Monte--Carlo simulation, 
and with which
we want to make the comparison. We have approximately, using eq. (49), the character
of the identity,
$$\bar\chi_{1,1}(q)={1-q-q^4+q^7+\ldots \over (1-q)(1-q^2)(1-q^3)\ldots },\e$$
and of the field $\phi$,
$$\bar\chi_{1,2}(q)={1-q^2-q^3+q^9+\ldots \over (1-q) (1-q^2) (1-q^3)\ldots},\e$$
where we defined, as usual,
$$q=\exp(2 \pi i\tau),\e$$
and were, for the comparison,
$\tau=i r/l$, following eq. (12). The square of one of these two quatities, 
we expect to recover 
in the Monte--Carlo simulation, according to eq. (12).

Now, in the conformal limit, $m\rarrow 0$, we can, actually, solve exactly the Bethe ansatz
for the SLYM system, eq. (5).
Inspecting the scattering matrix in this case, $\delta_{AA}(\beta)$, 
we see from eq. (26) that 
$$S_{AA}(\beta)=e^{i\delta_{AA}(\beta)},\e$$
and that for $\beta=0$, $\delta_{AA}(\beta)=\pi$ and that for $\beta>>1$, 
$\delta_{AA}(\beta)=0$.
This means that in eq. (5), since we can assume that if $\beta_i-\beta_j$ 
is not zero, it is
very big, so if $n_i\neq n_j$ then the Bethe equation for $i$ and $j$ decouples. 
This implies
that we need to solve the partition function only for $n_i$'s which are all
the same. In this case, we simply need to take $n_i\rarrow n_i-1/2$ if the
number of particles is even, and to leave it unchanged, $n_i$, if the number 
of particles is odd.

We conclude that the partition function for SLYM is:
$$Z(r/l)=\prod_{s=1}^\infty \left( 1+q^s+q^{2 s-1}+q^{3 s}+q^{4 s-2}+q^{5 s}+
q^{6 s-3}+\ldots\right)^2.\e$$
(We omited the zero mode, which gives an infinite factor, and as explained before,
it can be controlled by adding a chemical potential for the zero mode.)
It is immediately observed that this partition function, eq. (71), is different from the 
square of the characters, eqs. (67--68). We conclude that the scattering matrix, eq. (26)
needs to be modified by a ``Z-factor'' to make it correct for the SLYM system.
We conjecture that it needs to be multiplied by the solution of the Yang Baxter
equation based on the Hard Hexagon model, $IRF((G_2)_1,\phi,\phi)$ \r\Long. Further
work on this is required.

We hope that the Monte--Carlo simulation approach to the thermodynamic
Bethe ansatz can be a valuable tool in the investigation of integrable systems,
and as demonstrated here it can lead to very precise results.

\overfullrule=0pt

%
%
\setbox\strutbox=\hbox{\vrule height12pt depth8pt width0pt}
\midinsert
\line{\hfill Table 1.\hfill}
\line{\hfill Free Fermion Results, $r/l\leq 1$.\hfill}
\vskip15pt
\line{\hfill
\vbox{\offinterlineskip\hrule
\halign{&\vrule#&
  \strut\quad\hfil#\hfil\quad\cr
height2pt&\omit&&\omit&&\omit&&\omit&&\omit&&\omit&\cr
&$r/l$&&$\bar E_{\rm mc}$&&$N_{\rm mc}$&&$c_{\rm mc}$&&$\bar E_{\rm calc}$&&$N_{\rm calc}$&\cr
\noalign{\hrule}
height2pt&\omit&&\omit&&\omit&&\omit&&\omit&&\omit&\cr
& 0.05&&5.23606&&4.39815&&0.49876&&5.24908&&4.40616&\cr
&0.1&&2.64036&&2.19051&&0.49928&&2.64417&&2.19318&\cr
&0.15&&1.78325&&1.45019&&0.499623&&1.7846&&1.451&\cr
&0.2&&1.36164&&1.0766&&0.500103&&1.36135&&1.07633&\cr
&0.25&&1.11179&&0.847974&&0.499617&&1.11256&&0.848415&\cr
&0.3&&0.951349&&0.694035&&0.500076&&0.950611&&0.693539&\cr
&0.35&&0.838011&&0.580287&&0.499037&&0.837359&&0.580059&\cr
&0.4&&0.753898&&0.492667&&0.496496&&0.753123&&0.492233&\cr
&0.45&&0.686255&&0.421321&&0.490472&&0.686626&&0.421479&\cr
&0.5&&0.630722&&0.362648&&0.481836&&0.631075&&0.362824&\cr
&0.55&&0.583235&&0.313676&&0.470359&&0.582343&&0.313237&\cr
&0.6&&0.536727&&0.270202&&0.452238&&0.537975&&0.270783&\cr
&0.65&&0.496334&&0.234034&&0.433148&&0.496578&&0.234155&\cr
&0.7&&0.458426&&0.202888&&0.411322&&0.457424&&0.202418&\cr
&0.75&&0.419849&&0.174717&&0.384889&&0.420183&&0.174866&\cr
&0.8&&0.385041&&0.151055&&0.358719&&0.384752&&0.150935&\cr
&0.85&&0.351453&&0.13026&&0.331229&&0.351144&&0.130158&\cr
&0.9&&0.319961&&0.112321&&0.303853&&0.319412&&0.112135&\cr
&0.95&&0.289654&&0.0965256&&0.276236&&0.289617&&0.0965201&\cr
&1.&&0.262427&&0.0832056&&0.250599&&0.261799&&0.0830094&\cr 
height2pt&\omit&&\omit&&\omit&&\omit&&\omit&&\omit&\cr
}\hrule}
\hfill}
\setbox\strutbox=\hbox{\vrule height8.5pt depth3.5pt width 0pt}
\vskip20pt\endinsert

\overfullrule=0pt

%
%
\setbox\strutbox=\hbox{\vrule height12pt depth8pt width0pt}
\midinsert
\line{\hfill Table 2.\hfill}
\line{\hfill Free Fermion Results, $r/l\geq 1$.\hfill}
\vskip15pt
\line{\hfill
\vbox{\offinterlineskip\hrule
\halign{&\vrule#&
  \strut\quad\hfil#\hfil\quad\cr
height2pt&\omit&&\omit&&\omit&&\omit&&\omit&&\omit&\cr
&$r/l$&&$\bar E_{\rm mc}$&&$N_{\rm mc}$&&$c_{\rm mc}$&&$\bar E_{\rm calc}$&&$N_{\rm calc}$&\cr
\noalign{\hrule}
height2pt&\omit&&\omit&&\omit&&\omit&&\omit&&\omit&\cr
&1.&&0.262017&&0.0830768&&0.250208&&0.261799&&0.0830094&\cr
&1.05263&&0.234377&&0.0706784&&0.22352&&0.234671&&0.0707662&\cr
&1.11111&&0.206624&&0.0590791&&0.196221&&0.207096&&0.0592151&\cr
&1.17647&&0.179296&&0.0484496&&0.168978&&0.179385&&0.0484738&\cr
&1.25&&0.151874&&0.0386435&&0.141491&&0.151936&&0.0386596&\cr
&1.33333&&0.124918&&0.0298078&&0.114516&&0.125233&&0.0298832&\cr
&1.42857&&0.100163&&0.0223123&&0.089871&&0.0998352&&0.0222393&\cr
&1.53846&&0.0764025&&0.0158057&&0.0666761&&0.0763599&&0.015797&\cr
&1.66667&&0.0555191&&0.0106026&&0.0467796&&0.0554371&&0.0105871&\cr
&1.81818&&0.0377313&&0.00660547&&0.0304291&&0.0376456&&0.00659048&\cr
&2.&&0.0233141&&0.0037105&&0.0178107&&0.0234235&&0.00372794&\cr 
height2pt&\omit&&\omit&&\omit&&\omit&&\omit&&\omit&\cr
}\hrule}
\hfill}
\setbox\strutbox=\hbox{\vrule height8.5pt depth3.5pt width 0pt}
\vskip20pt\endinsert

\overfullrule=0pt

%
%
\setbox\strutbox=\hbox{\vrule height12pt depth8pt width0pt}
\midinsert
\line{\hfill Table 3.\hfill}
\line{\hfill Free Boson Results, $r/l\leq1$.\hfill}
\vskip15pt
\line{\hfill
\vbox{\offinterlineskip\hrule
\halign{&\vrule#&
  \strut\quad\hfil#\hfil\quad\cr
height2pt&\omit&&\omit&&\omit&&\omit&&\omit&&\omit&\cr
&$r/l$&&$\bar E_{\rm mc}$&&$N_{\rm mc}$&&$c_{\rm mc}$&&$\bar E_{\rm calc}$&&$N_{\rm calc}$&\cr
\noalign{\hrule}
height2pt&\omit&&\omit&&\omit&&\omit&&\omit&&\omit&\cr
&0.1&&4.29141&&3.80843&&0.906032&&4.28835&&3.80781&\cr
&0.15&&2.56987&&1.83794&&0.860102&&2.5692&&1.8375&\cr
&0.2&&1.72245&&1.03771&&0.816261&&1.72271&&1.03759&\cr
&0.25&&1.22361&&0.637259&&0.774554&&1.22529&&0.638052&\cr
&0.3&&0.903667&&0.414006&&0.737837&&0.902409&&0.41351&\cr
&0.35&&0.67905&&0.277354&&0.702127&&0.679256&&0.277446&\cr
&0.4&&0.517753&&0.190424&&0.670264&&0.518432&&0.190651&\cr
&0.45&&0.399507&&0.1334&&0.642885&&0.399148&&0.133236&\cr
&0.5&&0.308222&&0.0940281&&0.617436&&0.308909&&0.094242&\cr
&0.55&&0.239491&&0.0671612&&0.596375&&0.239727&&0.067244&\cr
&0.6&&0.186348&&0.0483229&&0.578307&&0.186231&&0.0482849&\cr
&0.65&&0.144824&&0.0348669&&0.562735&&0.14465&&0.0348313&\cr
&0.7&&0.112077&&0.025171&&0.549186&&0.112247&&0.0252108&\cr
&0.75&&0.0866057&&0.0182176&&0.537761&&0.086975&&0.0182923&\cr
&0.8&&0.0676613&&0.0133725&&0.528855&&0.0672725&&0.0132962&\cr
&0.85&&0.0519823&&0.00968116&&0.520219&&0.0519312&&0.00967727&\cr
&0.9&&0.0398549&&0.00702419&&0.512675&&0.0400067&&0.00705006&\cr
&0.95&&0.0308441&&0.00514874&&0.506253&&0.0307573&&0.00513968&\cr
&1.&&0.0235342&&0.00373636&&0.499938&&0.0235988&&0.00374886&\cr
height2pt&\omit&&\omit&&\omit&&\omit&&\omit&&\omit&\cr
}\hrule}
\hfill}
\setbox\strutbox=\hbox{\vrule height8.5pt depth3.5pt width 0pt}
\vskip20pt\endinsert

\overfullrule=0pt

%
%
\setbox\strutbox=\hbox{\vrule height12pt depth8pt width0pt}
\midinsert
\line{\hfill Table 4.\hfill}
\line{\hfill Free Boson Results, $r/l\geq1$.\hfill}
\vskip15pt
\line{\hfill
\vbox{\offinterlineskip\hrule
\halign{&\vrule#&
  \strut\quad\hfil#\hfil\quad\cr
height2pt&\omit&&\omit&&\omit&&\omit&&\omit&&\omit&\cr
&$r/l$&&$\bar E_{\rm mc}$&&$N_{\rm mc}$&&$c_{\rm mc}$&&$\bar E_{\rm calc}$&&$N_{\rm calc}$&\cr
\noalign{\hrule}
height2pt&\omit&&\omit&&\omit&&\omit&&\omit&&\omit&\cr
&1.&&0.0234887&&0.00373087&&0.499895&&0.0235988&&0.00374886&\cr
&1.05263&&0.0177955&&0.00268717&&0.493809&&0.0178181&&0.00269044&\cr
&1.11111&&0.0130254&&0.00186362&&0.487197&&0.0130086&&0.00186161&\cr
&1.17647&&0.00916301&&0.00123849&&0.479864&&0.00912634&&0.00123387&\cr
&1.25&&0.00611186&&0.00077747&&0.471513&&0.00610499&&0.00077701&\cr
&1.33333&&0.00386213&&0.00046036&&0.461907&&0.00385577&&0.000460143&\cr
&1.42857&&0.00225915&&0.00025103&&0.450652&&0.00227017&&0.000252884&\cr
&1.53846&&0.00125341&&0.00012921&&0.437441&&0.00122546&&0.000126766&\cr
&1.66667&&0.000561796&&0.00005319&&0.421766&&0.000593164&&0.0000566413&\cr
&1.81818&&0.000261345&&0.00002242&&0.403444&&0.000249743&&0.0000218611&\cr
&2.&&0.000104389&&0.00000785&&0.382052&&0.0000876474&&0.00000697&\cr
height2pt&\omit&&\omit&&\omit&&\omit&&\omit&&\omit&\cr
}\hrule}
\hfill}
\setbox\strutbox=\hbox{\vrule height8.5pt depth3.5pt width 0pt}
\vskip20pt\endinsert
\ack
I thank A. Babichenko, A. Belavin and S. Pal for helpful discussions.

\refout

\Appendix{}

{\bf Calculating the average energy and number of particles.}

Enclosed is a Mathematica program which calculates the partition
function $ff[a,l,mu]$, for one free fermion in a box of length $l$, 
the inverse
temperature $a$ and a chemical potential $mu$, using eq. (21). Then 
it calculates
the average energy $ee[a,l]$ and the average number of particles,
$nn[a,l]$. For free bosons we use the function $ffb[a,l,mu]$ listed below too, which gives
the partition function of one free boson, without the zero mode, and with
a chemical potential $mu$.

\obeylines

ff[a\_ ,l\_ ,mu\_ ]:=Product[(1+Exp[-mu-2 Pi a/l (n + 1/2)]) \^{} 2,$\{$n,0,60$\}$]
ee[a\_ , l \_] := -x D[Log[ff[x,l,0.]],x] /. x $->$ a
nn[a\_ ,l\_ ]:=-D[Log[ff[a,l,mu]],mu] /. mu $->$ 0.
Print[ee[0.05,1]," ",nn[0.05,1]]
5.24908 4.40616
ffb[a\_,l\_,mu\_]:=Product[(1-Exp[-mu-2 Pi a/l n])\^{}2,$\{$n,1,60$\}$]
\bye

%% file: mydef.tex
\def\e{\adveq\eqno{\rm (\chapterlabel\the\equanumber)}}

\def\adveq{\global\advance\equanumber by 1}
\def\myeq{{\rm \chapterlabel\the\equanumber}}
\def\rarrow{\rightarrow}

\def\semidirect{\mathrel{\raise0.04cm\hbox{${\scriptscriptstyle |\!}$
\hskip-0.175cm}\times}}

\def\mod{\mathop{\rm mod}\nolimits}

\def\ref#1{$^{[#1]}$}

\def\pr#1{#1^\prime}
 
\def\r#1{$[\rm#1]$} 
\def\twidle{\tilde}

\def\Tr{\mathop{\rm Tr}\limits}
\def\e{\adveq\eqno{\rm (\chapterlabel\the\equanumber)}}

\def\adveq{\global\advance\equanumber by 1}
\def\myeq{{\rm \chapterlabel\the\equanumber}}
\def\rarrow{\rightarrow}

\def\semidirect{\mathrel{\raise0.04cm\hbox{${\scriptscriptstyle |\!}$
\hskip-0.175cm}\times}}

\def\mod{\mathop{\rm mod}\nolimits}

\def\ref#1{$^{[#1]}$}

\def\pr#1{#1^\prime}
 
\def\r#1{$[\rm#1]$} 
\def\twidle{\tilde}

\def\Tr{\mathop{\rm Tr}\limits}